*Structural, magnetic, and electronic properties of GdTiO$_3$ Mott insulator thin films grown by pulsed laser deposition*

M. N. Grisolia[1], F. Y. Bruno[1], D. Sando[1§], H. J. Zhao[2,3], E. Jacquet[1], X. M. Chen[3], L. Bellaiche[2], A. Barthélémy[1] and M. Bibes[1]*

1. Unité Mixte de Physique, CNRS-Thales, 1 Av. Augustin Fresnel, Campus de l'Ecole Polytechnique, Palaiseau 91120, France, and Université Paris-Sud, 91405 Orsay, France
2. Physics Department and Institute for Nanoscience and Engineering, University of Arkansas, Fayetteville, Arkansas 72701, USA
3. Laboratory of Dielectric Materials, Department of Materials Science and Engineering, Zhejiang University, Hangzhou 310027, China

We report on the optimization process to synthesize epitaxial thin films of GdTiO$_3$ on SrLaGaO$_4$ substrates by pulsed laser deposition. Optimized films are free of impurity phases and are fully strained. They possess a magnetic Curie temperature T$_C$ = 31.8 K with a saturation magnetization of 4.2 $\mu_B$ per formula unit at 10 K. Transport measurements reveal an insulating response, as expected. Optical spectroscopy indicates a band gap of ~0.7 eV, comparable to the bulk value. Our work adds ferrimagnetic orthotitanates to the palette of perovskite materials for the design of emergent strongly correlated states at oxide interfaces using a versatile growth technique such as pulsed laser deposition.

§ now at Center for Correlated Electron Systems, Institute for Basic Science, Seoul National University, Seoul 151-747, Korea

* corresponding author : manuel.bibes@thalesgroup.com

---

Rare earth titanate perovskites RTiO$_3$ (where R is a trivalent rare earth ion such as Gd, Sm, Pr, …), have a Mott insulating character and rich magnetic phases arising from the coupling of the t$_{2g}$ orbitals and spin in the 3d$^1$ state through strong electronic correlations[1]. For smaller rare earths, the compounds are ferromagnetic and insulating, which is very unusual in simple perovskites, and thus offer interesting opportunities for oxide spintronics through e.g. spin filtering. Thin films of rare earth titanates may also allow the study of emergent concepts, such as correlated two-dimensional (2D) electron gases at their interfaces[2,3], or Mottronics systems[4,5].

In this work, we focus on thin films of ferrimagnetic and insulating GdTiO$_3$. In the bulk, GdTiO$_3$ crystallizes in an orthorhombic *Pbnm* structure (with parameters a=5.402 Å, b=5.697 Å, and c=7.68 Å) and presents a strong GdFeO$_3$-type distortion[1,6]. Its magnetic properties have been described as a Ti lattice of spin S=1/2 ordered ferromagnetically below 32 K and antiferromagnetically coupled to the ferromagnetically ordered Gd lattice (spin S=7/2), giving rise to a saturation magnetization of 6 $\mu_B$/f.u.[7] Here, we report on the growth of thin films of GdTiO$_3$ by pulsed laser deposition (PLD). Extensive structural characterization indicates that it is possible to obtain single phase films in appropriate deposition conditions. X-ray diffraction (XRD) is used to elucidate the structure obtained under optimum parameters. Samples grown in parasitic-phase-free conditions display magnetic and electronic characteristics comparable to values reported for $GdTiO_3$ thin films grown by molecular beam epitaxy (MBE) [8,9].

The growth of stoichiometric $RTiO_3$ thin films is particularly challenging due to the unstable valence of $Ti^{3+}$ ions in an oxygen-rich atmosphere. Typically, the pyrochlore $Gd_2Ti_2O_7$ phase is synthesized more easily than the perovskite $GdTiO_3$ (GTO), cf. Ref. [10]. In bulk form, GTO can only be prepared in strongly reduced atmosphere, e.g. in a mixture of hydrogen and argon[11]. Here, films were grown on (001)-oriented $SrLaGaO_4$ (SLGO) substrates with an in-plane lattice parameter $a_{ip}$ = 3.852 Å by PLD from a stoichiometric pyrochlore target using a Nd:yttrium aluminum garnet (YAG) laser at 355 nm. Prior to deposition, the substrate was cleaned up with acetone in an ultrasound bath for 5 min before being placed in an isopropanol bath for another 5 min. In the chamber, the SLGO crystal was heated by a resistive wire under an oxygen pressure of 0.4 mbar up to the deposition temperature. Then the pressure was lowered to the deposition value. The structure of the films was monitored during growth using in situ reflection high energy electron diffraction (RHEED). After growth the films were further characterized by XRD using a Panalytical Empyrean equipped with a Ge(220) monochromator and a PIXcel detector.

A series of films was grown at a constant fluence of 1.1 J/cm² while changing the temperature and oxygen pressure in order to establish the growth window of single phase GTO. After deposition the films were cooled down to room temperature at a constant rate in the growth $O_2$ partial pressure. Fig. 1a and Fig. 1b present XRD spectra for samples grown at constant temperature while varying pressure, and at constant pressure while varying temperature, respectively. We find that the pure pyrochlore ($Gd_2Ti_2O_7$) phase is stabilized at low temperature and low oxygen partial pressure, as highlighted by the vertical dotted red lines. Upon increasing pressure and temperature to $PO_2=10^{-6}$ mbar and T=650°C, the pure GTO phase is stabilized (vertical blue lines). Further lowering the temperature stabilizes spurious Magnéli phases like $Ti_6O_{11}$ (vertical green lines). Fig. 1d gathers XRD spectra displaying the whole variety of phases encountered during the optimization. The temperature-pressure phase diagram of Fig. 1e summarized these results. Fig. 1c presents XRD data of films grown at optimal pressure and temperature, while varying the laser fluence. All films are single phase and the out-of-plane parameter remains unchanged. The data thus suggest that, as the stoichiometry of GTO is concerned, pressure is the main driving parameter, with laser fluence and temperature playing a less prominent role.

In Fig. 1f, we present some characterization of the surface quality of our films at the end of the growth. As it can be observed, the RHEED patterns are still clearly visible displaying all the features of a 2D surface for a 15 nm-thick film. The top panel of figure 1f shows an atomic force microscopy (AFM) image of a film. Atomic steps and terraces are visible, on top of a more wavy structure. The root-mean-square roughness of such single phase GTO films is typically 1-2 nm.

Due to the strongly distorted orthorhombic unit cell of GTO, it is not straightforward to predict the most favorable growth orientation on $[001]_s$ SLGO (where the subscript s refers to the substrate). Two possibilities exist with either the $[001]_o$ or the $[110]_o$ GTO axis parallel to the substrate normal (here the o subscript refers to the orthorhombic unit cell of GTO). In Figure 2, we present the different growth orientations ($[110]_o$ and $[001]_o$ respectively) along with their epitaxial relationships and visualizations. If the film grows along [110]$_o$ (Fig 2a) the [-110]$_o$ and [001]$_o$ directions lie in the plane of the substrate (Fig. 2b and Fig. 2d) and in fully-strained films the lattice vector along these directions equal to 2a$_{ip}$. Although the lattice vector along the diagonal [-110]$_o$ is fixed to the substrate, the ratio a$^f$/b$^f$ (where the subscript f refers to film) is free to change since it can easily be accommodated by the tilting of the TiO$_6$ octahedra (Fig. 2a) and a consequent change of the angle α or β as represented in Fig. 2c. On the other hand, if the film grows along [001]$_o$ (Fig. 2e), the constraints are that a$^f$ and b$^f$ must be equal to $\sqrt{2}a_{ip}$ (Fig. 2h).

For both growth orientations, two different in plane strain values can be considered.. For the [110]$_o$ growth direction, one can define a strain sate along the c direction of the orthorhombic unit cells ([001]$_o$) as $\epsilon_1 = \frac{2a_{ip}-c}{c}$ and another strain state corresponding to the diagonal direction ([1-10]$_o$) $\epsilon_2 = \frac{2a_{ip}-(\sqrt{a^2+b^2})}{(\sqrt{a^2+b^2})}$. In the case of the [001]$_o$ oriented growth, one obtains the a and b parameters of the film lying in the plane but rotated by 45 degree with respect to the in-plane parameters of the substrate. This yields the following strain states: $\epsilon_1 = \frac{\sqrt{2}a_{ip}-a}{a}$ and $\epsilon_2 = \frac{\sqrt{2}a_{ip}-b}{b}$. Calculated in this way the average in-plane misfit strain is -3% for the [001]$_o$ oriented growth and -1% for [110]$_o$. Thus, to minimize elastic energy the film should grow with the [110]$_o$ axis parallel to the SLGO substrate normal [001]$_s$. Such an orientation has already been observed in *Pnma* systems such as CaTiO$_3$ and is referred to as *ab-ePnma*[12] for which the c axis lies in the plane while a and b are tilted out of the plane, as schematically shown in Fig. 2a and Fig. 2c.

To confirm these predictions, we performed reciprocal space mapping (RSM) around the (33-2)$_o$ peak of GdTiO$_3$ (which, without taking into account the distortion, may be referred to as the (-103)$_{pc}$ peak in pseudocubic notation) and the (-107)$_s$ peak of SLGO. The in-plane components corresponding to the peaks of GTO and SLGO in Fig. 3a coincide, indicating that the film is fully strained. RSMs around (200)$_o$ reflections measured with the sample at χ=45 degrees were also performed to determine the other lattice parameters of the film. As visible from Fig. 3b, two peaks are present, corresponding to two structural variants. These two peaks are indeed the superposition of domains rotated by 180 degrees with respect to each other around the <110>$_o$ direction. In Fig. 3c, we display a view of the two structural variants along the <001>$_o$ direction. Since χ is defined with respect to the substrate and a differs from b, both directions (<0-10>$_o$ and <100>$_o$) cannot simultaneously be perpendicular to the sample holder. The slight tilt will be denoted in what follows as κ and the lattice parameters extracted from Fig. 3b as $a^{\chi=45}_{[100]_o}$, $b^{\chi=45}_{[010]_o}$.

Assuming that the orientation of the unit cells is the one described in the diagram Fig. 2c, we can find all structural parameters by solving 7 equations simultaneously where c$_{out}$ is calculated from the 2θ-θ scan shown in Fig. 1a, and $a^{\chi=45}_{[100]_o}$, $b^{\chi=45}_{[010]_o}$ are both extracted from Fig. 3b.

$$\frac{d_{[110]_o}}{2} \sin \kappa = c_{out} \tag{1}$$

$$a^2 + b^2 - 2ab \cos \gamma = 4a_{ip}^2 \tag{2}$$

$$2 \tan^{-1}\left(\frac{b}{a}\right) = \kappa \tag{3}$$

$$\cos\left(\frac{\pi}{4} - \alpha\right) b = b^{\chi=45}_{[010]_o} \tag{4}$$

$$\cos\left(\frac{\pi}{4} - \beta\right) a = a^{\chi=45}_{[100]_o} \tag{5}$$

$$a^2 + b^2 + 2ab \cos \kappa = d_{[110]_o}^2 \quad (6)$$
$$\alpha + \beta = \pi - \gamma \quad (7)$$

Solving this system numerically yields the following values: a=5.47 Å, b=5.56 Å, $d_{[110]_o}$=7.85 Å, γ=88.8°, κ=90.3°. These values confirm that the film is grown in the [110]$_o$ orientation, lowering its symmetry from *Pbnm* to *P2$_1$/m* in order to accommodate the in plane strain (similar to the case of SrRuO$_3$ on SrTiO$_3$, see Ref. [13]).

Next, we characterized the physical properties of single phase GTO films grown at 1.1 J/cm², 660°C and 2.5 10$^{-6}$ mbar by transport, optical, and magnetic measurements. The electrical resistivity is shown in Fig. 4a. The resistivity increases with decreasing temperature, consistent with a Mott insulating behavior[8]. It is well-known that oxygen-rich rare earth titanate films are metallic[10], thus the insulating nature of our GTO films suggests that no major off-stoichiometry is present[11,14]. At room temperature the resistivity is 19.8 Ω cm, which is lower than that of bulk GTO (26 Ω.cm)[15] but higher than the values obtained in previous studies on thin films[8,9]. The inset in Fig. 4a shows the resistivity as a function of the inverse temperature (1/T), and a fit of the data using Arrhenius law. The activation energy is found to be 0.094 eV and is comparable to the value of 0.14 eV for thin films found in Ref. [8] but lower than the bulk value known to be 0.19-0.23 eV (Refs. [1,16]).

To determine the optical band gap of our GdTiO$_3$ thin films, we performed transmission spectroscopy using a Cary spectrometer over the spectral range 1000-2000 nm (1.24-0.62 eV). The absorption coefficient of the GTO layer, $\alpha$, was calculated from the transmission of the GTO sample, *T*, and the substrate, *T$_0$*, using the relation $\alpha = -\frac{1}{t} \ln\left(\frac{T}{T_0}\right)$, where *t* is the thickness of the GTO layer (determined by X-ray reflectometry). To determine the band gap, a Tauc plot of $(\alpha E)^2$ vs $E$ was constructed, and the linear region extrapolated to the $E$ axis (Fig. 4b), yielding an optical gap value of ~0.7 eV in good agreement previous results in thin films[8] and bulk[17].

Figure 4c shows the magnetic properties of the GTO films. Magnetization as a function of the magnetic field is hysteretic with a coercive field (~370 Oe) at 10 K, consistent with the expected ferrimagnetism of GTO (Ref. [1,7]). The saturation magnetization at 10 K is about 4.2 μ$_B$ per formula unit (f.u.). This value is lower than the bulk value(~ 6 μ$_B$/f.u.) but is comparable to the value found in [9] for films grown on (LaAlO$_3$)$_{0.3}$-(Sr$_2$AlTaO$_6$)$_{0.7}$ (LSAT). In order to correctly remove the paramagnetic background arising from impurities in the substrate, we determined the M(T) by measuring the remanent magnetization from M(H) hysteresis loops performed at different temperatures after field cooling. The result is shown in Fig. 4d. A classical mean field fit of the ferromagnetic cycle gives a T$_c$ of 31.8 K very close to the corresponding bulk value of 32-34 K (Refs. [1,16]).

To gain further insight into our results, we performed theoretical density functional (DFT) calculations. They were carried out with the Vienna *ab initio* Simulation Package (VASP) [18] using projected augmented wave (PAW) potentials. The generalized gradient approximation plus Hubbard U method (GGA+U) within the framework of Perdew-Burke-Ernzerhof revised for solids (PBEsol)[19] was selected for this work. The on-site Coulomb interactions and Hubbard U were selected to be 4.0 eV and 3.0 eV for Gd$^{3+}$ and Ti$^{3+}$, respectively. We used 18 valence electrons for Gd (4f$^7$5s$^2$5p$^6$5d$^1$6s$^2$), 4 valence electrons for Ti (3d$^3$4s$^1$) and 6 valence electrons for O (2s$^2$sp$^4$). The plane wave energy cut off and ionic relaxation Hellmann-Feynman force convergence criteria were selected as 500 eV and 0.005 eV/Å, respectively. The space group of the different structural phases was determined by the FINDSYM code[20]. The two possible growth orientations [001]$_o$ and [110]$_o$ were considered and simulated. Considering the SLGO substrate with in-plane lattice parameter of ~3.852 Å, our calculation indicates that the P2$_1$/m phase (ab-ePbnm) is more stable than the Pbnm phase (c-ePbnm). The calculated mean value of the a and b lattice parameters of P2$_1$/m phase on SLGO substrate is 5.549 Å. The calculation overestimates the mean value of a and b by ~0.6% compared to the experimental 5.515 Å. The calculation shows the c lattice parameter of 7.704 Å and γ angle of 87.9 degree (the atomic positions presented in Fig. 2 and Fig. 3c are the actual atomic positions from the computations). The simulated GdTiO$_3$ on SLGO substrate shows a total magnetic moment of 5.99 μ$_B$, and a partial magnetic moment (magnitude) for Gd$^{3+}$ and Ti$^{3+}$ of 6.96 μ$_B$ and 0.94 μ$_B$, respectively. The calculation overestimates the magnetization possibly since the spins are assumed to be collinear, *i.e.* no canting or frustration is considered. Consistent with the experimental results, the theory predicts that the most favorable growth direction of GTO on SLGO is [110]$_o$. Although the values of the lattice constants are slightly different, the results agree in showing the monoclinic rearrangement of the atoms in the P2$_1$/m space group. The theory allows the calculation of the Ti-O-Ti angles in all directions. In the [110]$_o$ oriented growth the Ti-O-Ti angle along both the a and c direction decreases from 143.7° (calculated bulk value) to 141.5° and 141.1° respectively. We thus argue that the biaxial stress imposed by the epitaxial strain may be responsible for the different behavior found in films compared to the bulk. The modification of the bandwidth (W defined as $W \propto \cos^2\Theta$ where Θ is the angle of the R-O-R bonds[21]) by epitaxial strain has profound consequences on the physical properties of strongly correlated electronic systems[22]. A heterogeneous decrease in the Ti-O-Ti angle thus results in a decrease of the bandwidth and yields a lower activation energy.

In summary, we have reported on the various phases that are stabilized when growing GdTiO$_3$ films by PLD as a function of temperature, pressure and fluence during the optimization process. In optimal conditions, we are able to grow [110]$_o$ single phase, fully strained GdTiO$_3$ films that are monoclinic, ferrimagnetic and insulating. The sensitivity of the GdTiO$_3$ system to extrinsic defects is discussed by comparing the experimental results to bulk values and theoretical predictions. This work opens the way to the exploration by PLD of oxide heterostructures based on rare earth titanates.

Acknowledgements

We acknowledge financial support from the Labex NanoSaclay project FIRET. H. J. Z. acknowledges the financial support of National Science Foundation of China under Grant Nos. 51332006. L.B. thanks the Department of Energy, Office of Basic Energy Sciences, under contract ER-46612 and NSF Grant No. DMR-1066158. Some computations were also made possible thanks to the MRI Grant No. 0722625, MRI-R2 Grant No. 0959124, and Grant No. 0918970 from NSF, and a Challenge grant from the Department of Defense.

Word count

2404 words + 3 single column figures (126 words each) + one double column figure (464 words) + 7 equations (16 words each) = 3355 words < 3500 limit

References

[1] H.D. Zhou and J.B. Goodenough, J. Phys.: Condens. Matter **17**, 7395 (2005).

[2] S. Okamoto and A.J. Millis, Nature **428**, 630 (2004).

[3] J. Biscaras, N. Bergeal, A. Kushwaha, T. Wolf, A. Rastogi, R.C. Budhani, and J. Lesueur, Nature Commun. **1**, 89 (2010).

[4] M. Nakano, K. Shibuya, D. Okuyama, T. Hatano, S. Ono, M. Kawasaki, Y. Iwasa, and Y. Tokura, Nature **487**, 459 (2012).

[5] H. Yamada, M. Marinova, P. Altuntas, A. Crassous, L. Bégon-Lours, S. Fusil, E. Jacquet, V. Garcia, K. Bouzehouane, A. Gloter, J.E. Villegas, A. Barthélémy, and M. Bibes, Sci. Rep. **3**, 2834 (2013).

[6] A. Komarek, H. Roth, M. Cwik, W.-D. Stein, J. Baier, M. Kriener, F. Bourée, T. Lorenz, and M. Braden, Phys. Rev. B **75**, 224402 (2007).

[7] C.W. Turner and J.E. Greedan, J. Solid State Chem. **34**, 207 (1980).

[8] P. Moetakef, D.G. Ouellette, J.Y. Zhang, T.A. Cain, S.J. Allen, and S. Stemmer, J. Cryst. Growth **355**, 166 (2012).

[9] P. Moetakef, J.Y. Zhang, S. Raghavan, A.P. Kajdos, and S. Stemmer, J. Vac. Sci. Technol. A **31**, 041503 (2013).

[10] A. Ohtomo, D.A. Muller, J.L. Grazul, and H.Y. Hwang, Appl. Phys. Lett. **80**, 3922 (2002).

[11] M. Heinrich, H.-A. Krug von Nidda, V. Fritsch, and A. Loidl, Phys. Rev. B **63**, 193103 (2001).

[12] C.-J. Eklund, C. Fennie, and K. Rabe, Phys. Rev. B **79**, 1 (2009).

[13] Q. Gan, R. A. Rao, C.B. Eom, L. Wu, and F. Tsui, J. Appl. Phys. **85**, 5297 (1999).

[14] G. Amow, N.P. Raju, and J.E. Greedan, J. Solid State Chem. **155**, 177 (2000).

[15] M. Reedyk, D.A. Crandles, M. Cardona, J.D. Garrett, and J.E. Greedan, Phys. Rev. B **55**, 1442 (1997).

[16] J.E. Greedan, J. Less-Common Met. **111**, 335 (1985).

[17] D.A. Crandles, T. Timusk, J.D. Garrett, and J.E. Greedan, Physica C **201**, 407 (1992).

[18] G. Kresse, Phys. Rev. B **59**, 1758 (1999).

[19] J.P. Perdew, K. Burke, and M. Ernzerhof, Phys. Rev. Lett. **77**, 3865 (1996).

[20] H.T. Stokes and D.M. Hatch, J. Appl. Crystallogr. **38**, 237 (2005).

[21] M. Imada, A. Fujimori, and Y. Tokura, Rev. Mod. Phys. **70**, 1039 (1998).

[22] F.Y. Bruno, K.Z. Rushchanskii, S. Valencia, Y. Dumont, C. Carrétéro, E. Jacquet, R. Abrudan, S. Blügel, M. Ležaić, M. Bibes, and A. Barthélémy, Phys. Rev. B **88**, 195108 (2013).

Figure captions

Figure 1: XRD diagrams of 15 nm thick GTO films grown at 1.1 J/cm² and (a) constant temperature and varying pressure, and (b) constant pressure and varying temperature. (c) XRD diagrams of films grown at the same pressure and temperature and different fluence values. (d) XRD diagrams of the different phases identified during the growth optimization. (e) Phase diagram summarizing the growth optimization as a function of pressure and temperature for a fluence of 1.1 J/cm². (f) 2×1 µm² atomic force microscopy image of a 15 nm-thick film and RHEED images of the SLGO substrate before and after deposition of a GTO film, at 0 deg and 45 deg azimuthal angles from the [100] direction of $GdTiO_3$.

Figure 2: Sketches of the GTO orthorhombic unit cell growing onto $[001]_s$-oriented SLGO in the $[110]_o$ orientation (a-d) or the $[001]_o$ orientation (e-h). (a) and (e) are views in three dimensions. (b) and (f) are projections on the $[100]_s$ direction. (c) and (g) are projections on the $[010]_s$ direction. (d) and (h) are views from the top.

Figure 3: (a) Reciprocal space map near the $[-107]_s$ peak of SLGO and of $[33-2]_o$ of GTO for a film grown at 0.8 J/cm², 700°C and 2 $10^{-6}$ mbar. (b) Reciprocal space map near the $<200>_o$ peak at $\chi=45$ degrees. (c) Sketches displaying the projections of the two variants oriented at 180° with respect to each other at $\chi=45$ degrees along $<001>_o$.

Figure 4: (a) Electrical resistivity as a function of temperature of a 15 nm GTO film grown at 1.1 J/cm², 2.5 $10^{-6}$ mbar and T=660 °C. Inset: same plotted against the inverse temperature; an activated behavior characteristic of Mott insulating is used to fit the data. (b) UV spectroscopy data performed on the same sample at room temperature. (c) M(H) hysteresis loop and (d) M(T) for a 15 nm-thick film grown in the same conditions. In (d) the width of the red dots corresponds to the error bars in the measurements; the blue line is a fit to the data in a classical mean field model.

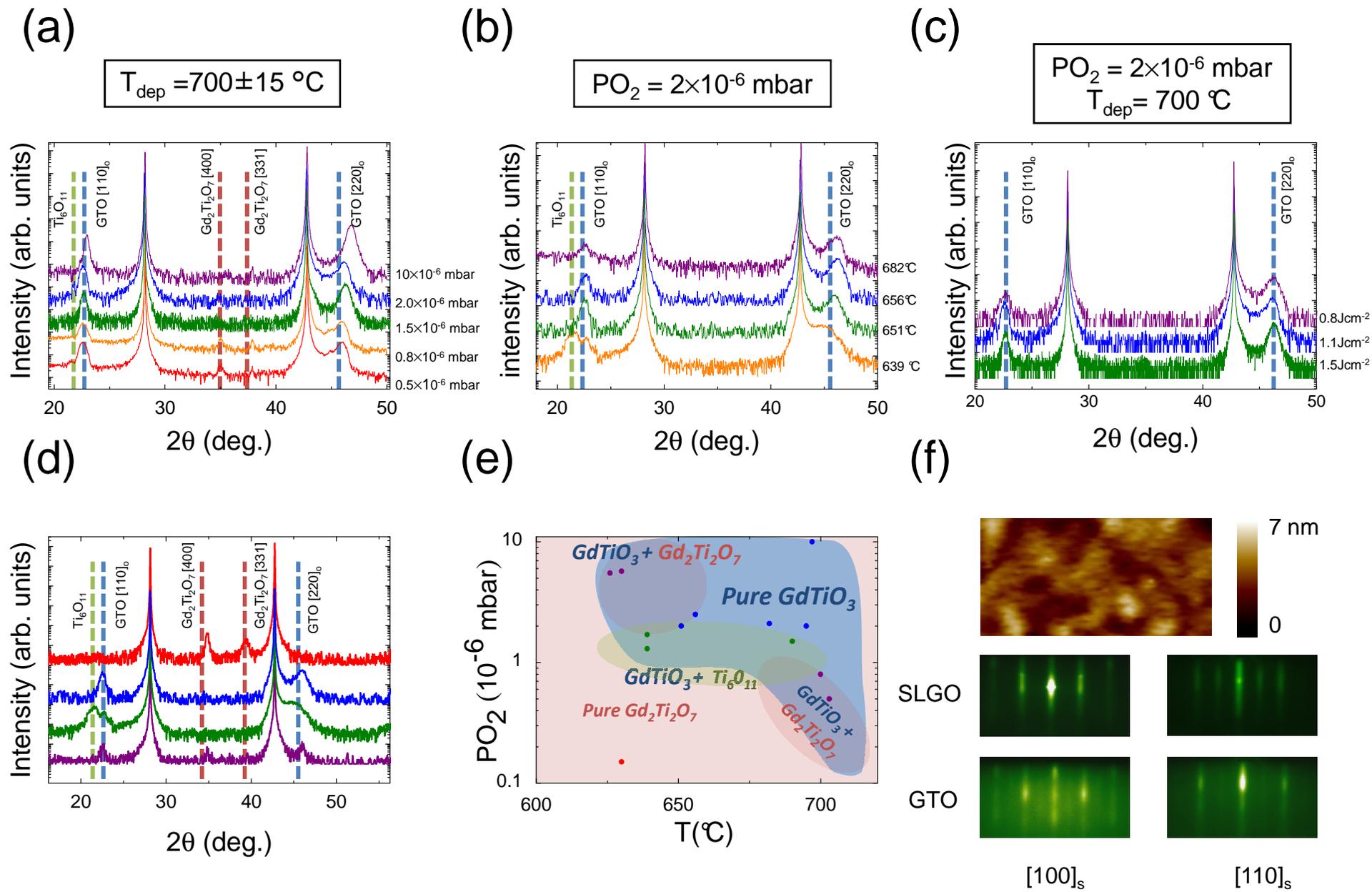

Fig 1. Grisolia et al (double column)

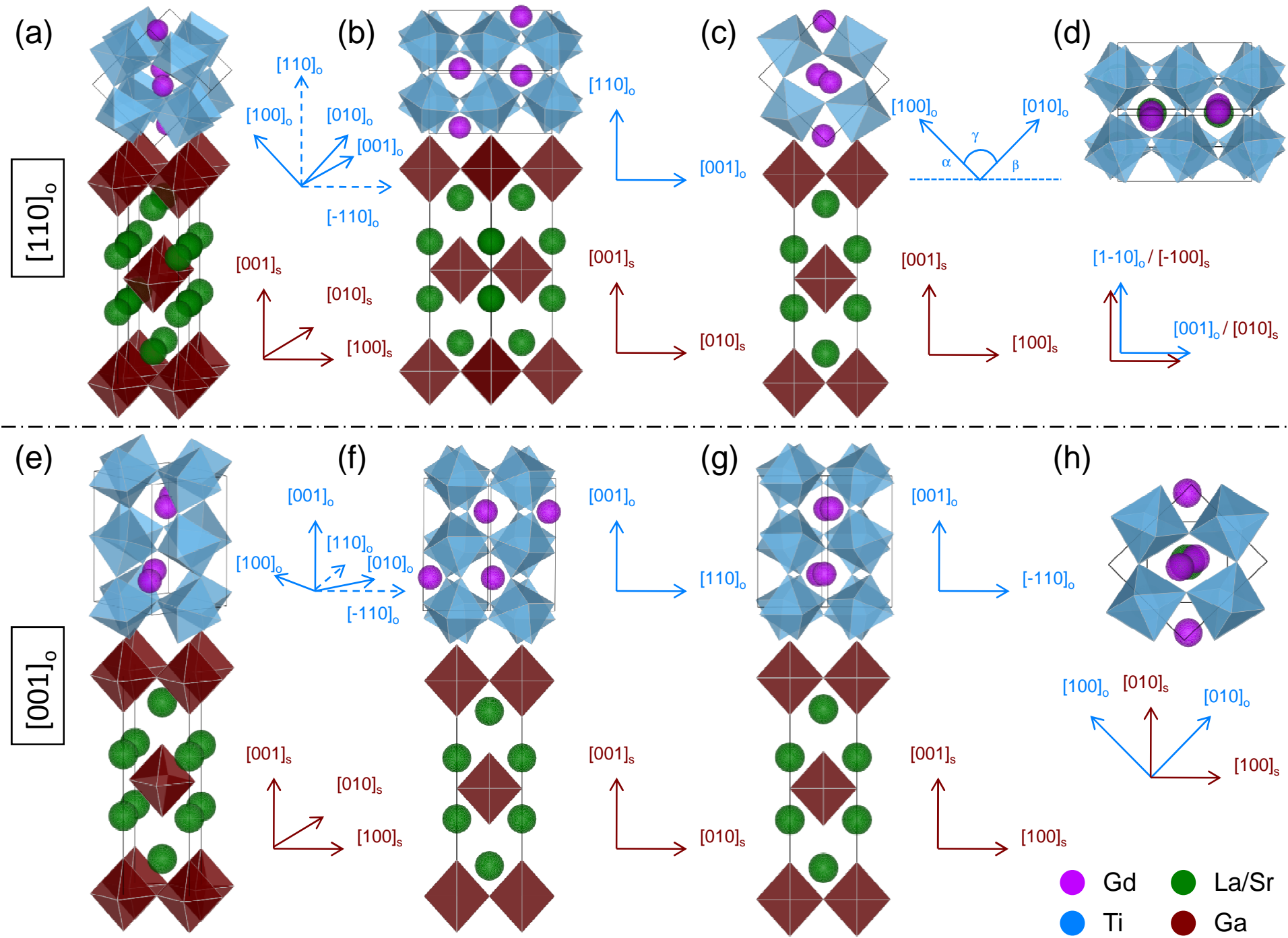

Fig. 2. Grisolia et al (double column)

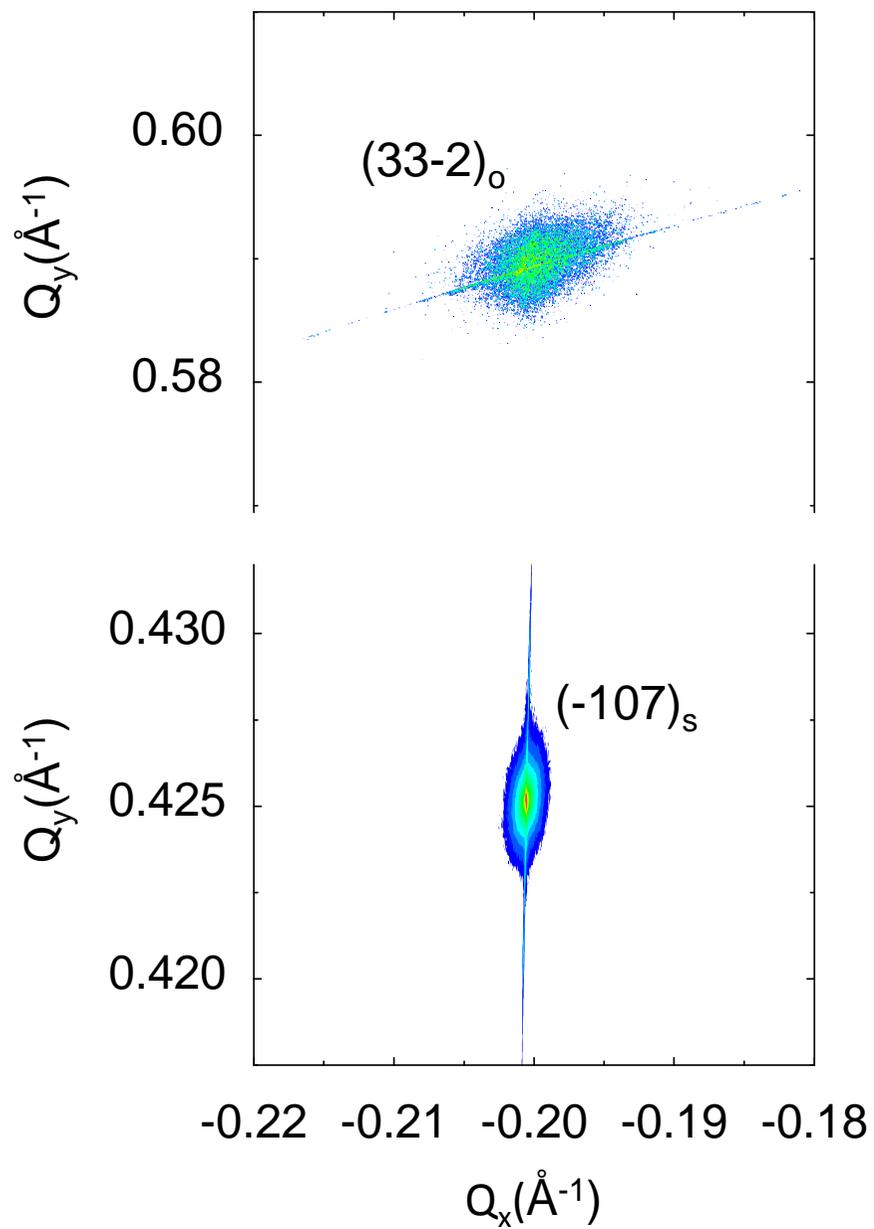
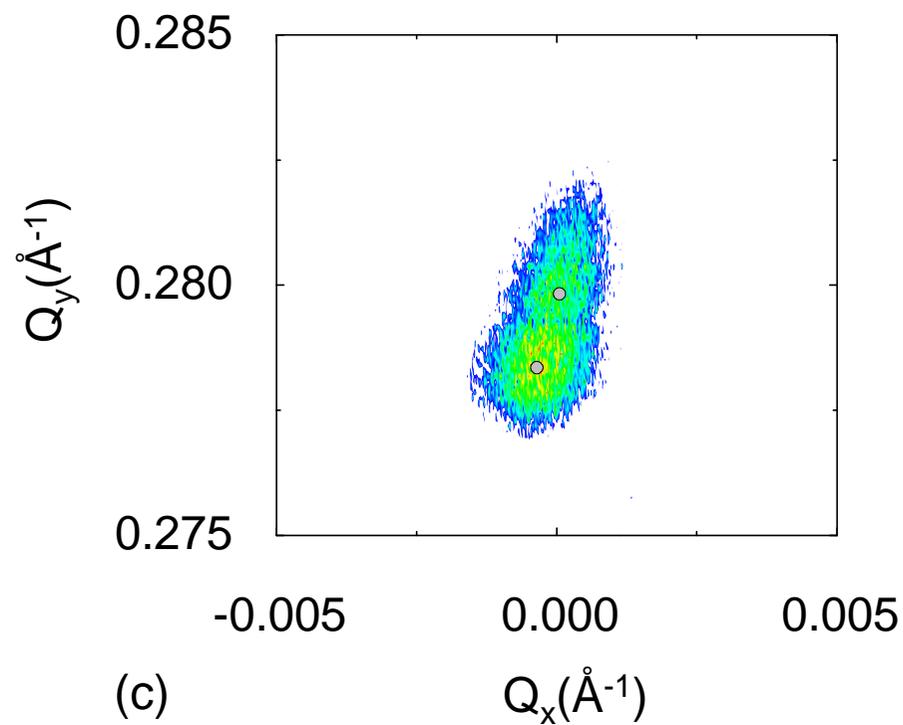
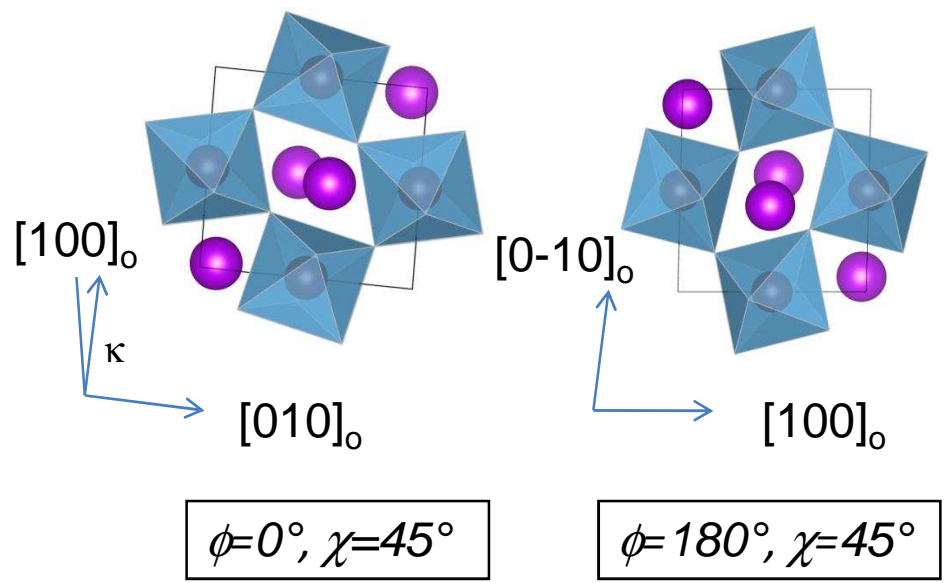

Fig. 3. Grisolia et al (single column)

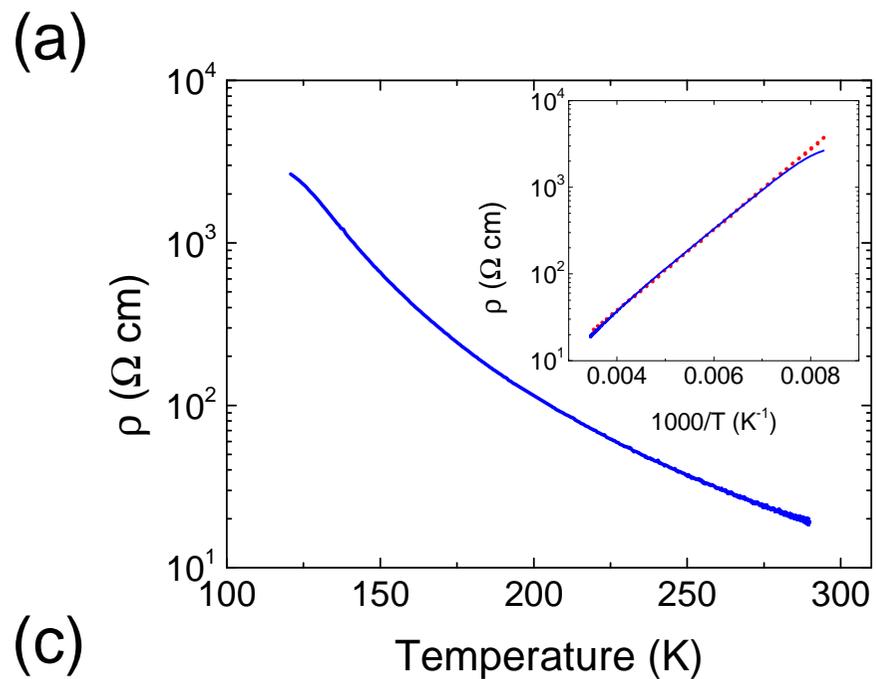
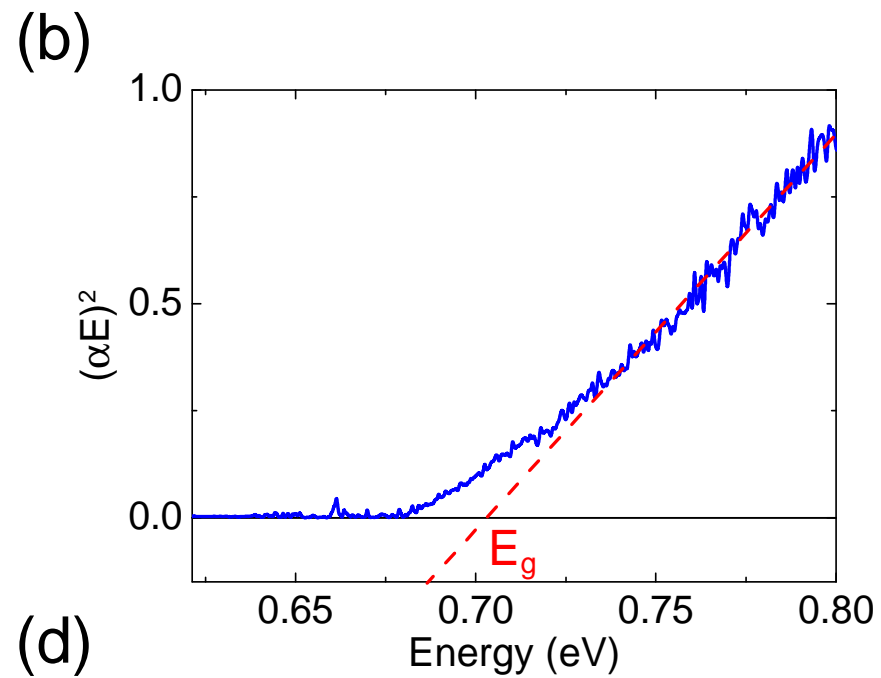
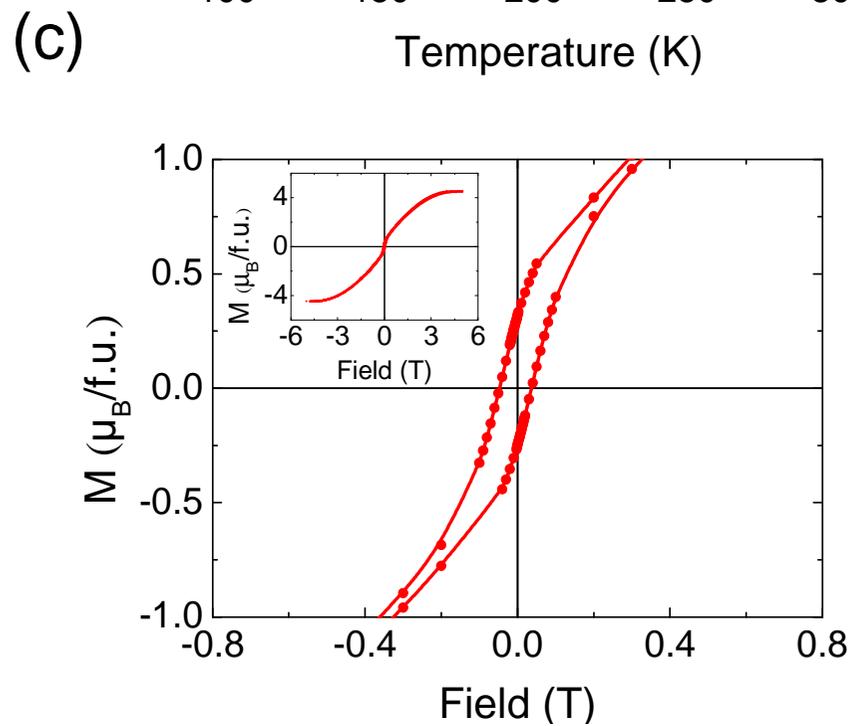
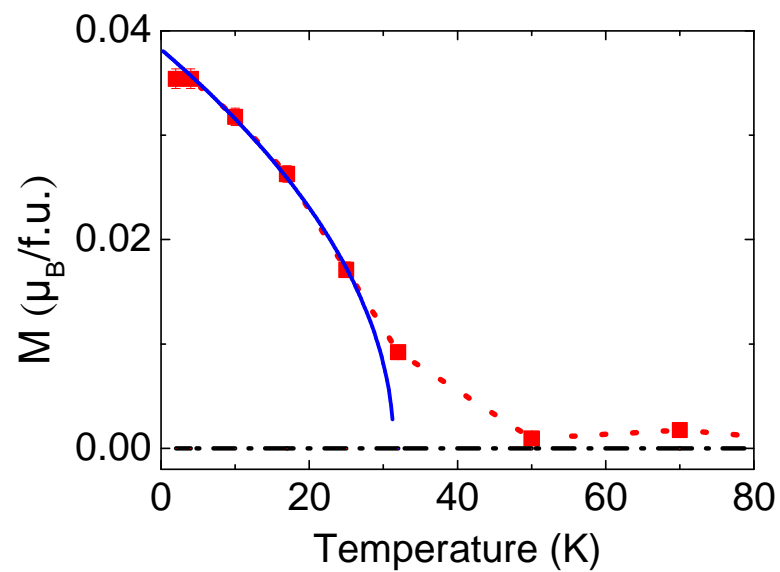

Fig. 4. Grisolia et al (single column)